\documentclass[aps,pra,twocolumn,10pt]{revtex4-2}
\usepackage[T1]{fontenc}
\usepackage{bm}
\usepackage{amsmath}
\usepackage{amssymb}
\usepackage{slashed}
\usepackage{float}
\allowdisplaybreaks
\usepackage{dcolumn}
\newcolumntype{d}[1]{D{.}{.}{#1}}
\newcommand*{\centt}[1]{\multicolumn{1}{c}{#1}}

\begin{document}

\title{Magnetic shielding in the atomic hydrogen anion}
\author{Tymon Kilich}
\email{Tymon.Kilich@fuw.edu.pl}
\author{Krzysztof Pachucki}
\email{Krzysztof.Pachucki@fuw.edu.pl}
\affiliation{Faculty of Physics, University of Warsaw,
             Pasteura 5, 02-093 Warsaw, Poland}
    
\begin{abstract}
The atomic hydrogen anion H$^-$ is the lightest stable anion and its bound states and resonances are well studied in the literature. 
Due to the planned comparison of the bare antiproton to H$^-$ in a Penning trap, 
we study the magnetic shielding of H$^-$ using the nonrelativistic quantum electrodynamics theory, 
by accurately calculating the non-relativistic shielding,
as well as finite nuclear mass,
relativistic,
and partially QED corrections. 
We find that the finite nuclear mass correction is quite significant in H$^-$ contributing about $0.1\%$ of the total shielding,
which is more than twice as much as the relativistic correction. 
Our final result for the shielding constant has a nine-parts-per-trillion accuracy and
paves the way for direct comparison of the antiproton-to-proton magnetic moments.

\end{abstract}

\maketitle

\section{Introduction}
The simplest and lightest of anions -- the atomic hydrogen anion -- has been comprehensively studied theoretically in the past, 
starting from the theoretical prediction of the existence of the ground state by Bethe~\cite{Bethe1929} subsequently confirmed by the calculations of Hylleraas~\cite{Hylleraas1930}. 
Many studies investigate the electronic structure of the ground state of H$^-$~\cite{Thakkar1994, Drake2002, Harris2005, Frolov2015} with the most accurate calculations reaching more than 30 significant digits of the non-relativistic energy~\cite{Nakatsuji2007, Korobov2018, Drake2025}. Although the nuclear magnetic shielding of H$^-$ was previously calculated using quantum-chemical methods \cite{Boucekkine1988, Vaara2003}, no estimate of uncertainty is given in those studies and no recoil and relativistic corrections were taken into account. In this work, we obtain a highly accurate value of the nuclear magnetic shielding in the H$^-$ ion,
by calculation of all important finite nuclear mass, relativistic and quantum electrodynamic (QED) corrections.

Highly accurate calculations of nuclear magnetic shielding using the nonrelativistic QED (NRQED) theory have recently been done for the hydrogen atom and $^3$He cation~\cite{Pachucki2023_PRA} as well as the helium atom~\cite{Rudzinski2009_JCP, Wehrli2021_PRL, Wehrli2022_PRA}, while the combined NRQED and all-order approach was used to obtain the values of the shielding constants for helium-like cations~\cite{Yerokhin2024_PRA}. Combining the measurement of the $^3$He$^+$ ion \cite{Schneider2022} together with the calculation of the magnetic shielding \cite{Wehrli2021_PRL, Wehrli2022_PRA, Pachucki2023_PRA} has allowed for the most accurate determination of the magnetic moment of helion 
and of the whole $^3$He atom. This result is important for the development of a new absolute magnetometry standard.

The motivation for the present calculation, in addition to the lack of an accurate value of the magnetic shielding constant for H$^-$ in the literature, comes from the progress in the experimental measurements of magnetic moments of proton~\cite{Schneider2022b} and antiproton~\cite{Smorra2017} with parts-per-billion accuracy as well as the antiproton-to-proton charge–mass ratio experiments~\cite{Borchert2022}, where H$^-$ serves as an excellent negatively charged proxy for the antiproton. The new generation of experiments using a multi-Penning-trap system of the BASE collaboration~\cite{Latacz2025} paves the way for determining magnetic moments of the proton and antiproton with a fractional accuracy of less than 100 ppt. The experimental setup should allow to perform a measurement of the magnetic moment of H$^-$ with respect to that of the antiproton with even better  accuracy~\cite{PrivCommSU_2025}.

\section{Theoretical method}

According to Ramsey's theory \cite{Ramsey1950}, the coupling of the nuclear magnetic moment $\vec\mu$ with the homogeneous magnetic field $\vec B$ is modified by the presence of atomic electrons
\begin{equation}
\delta H = -\vec\mu\cdot\vec B\,(1-\sigma), \label{eq:shielding_definition}
\end{equation}
where $\sigma$ is a dimensionless shielding constant. In the low-$Z$ atomic systems, the shielding constant
is represented as a double expansion in $\alpha$ and the electron-to-nucleus mass ratio $m/M$. The expansion in $\alpha$ takes the form
\begin{equation}
\sigma = \alpha^2 \sigma^{(2)} + \alpha^4 \sigma^{(4)} 
 + \alpha^5 \sigma^{(5)} + \ldots\,.   
\end{equation}
The expansion terms $\sigma^{(n)}$ are subsequently the non-relativistic shielding,
the relativistic, the leading QED, and the higher-order QED corrections to the shielding. 
They can be further expanded in the electron-nucleus mass ratio
\begin{align}
\sigma^{(n)} =&\  \sigma^{(n,0)} + \sigma^{(n,1)} + \sigma^{(n,2)} +\ldots
\end{align}

Detailed derivations of the  $\sigma^{(n,0)}$ terms, up to $n=5$, are presented in the previous works \cite{Pachucki2008_PRA, Rudzinski2009_JCP, Wehrli2022_PRA}, 
while the finite nuclear mass correction $\sigma^{(2,1)}$ was derived by one of us (KP) in Ref.~\cite{Pachucki2008_PRA} and later corrected in Ref.~\cite{Pachucki2023_PRA}.
Therefore, we present only the derived formulas, in a.u., used in the numerical evaluation of the nuclear magnetic shielding of H$^-$.

The leading-order nonrelativistic and nonrecoil $\sigma^{(2,0)}$ contribution to the nuclear magnetic shielding of a closed-shell, spherically symmetric atomic system was first derived by Lamb~\cite{Lamb1941} and later generalized to diatomic molecules by Ramsey~\cite{Ramsey1950}. In our notation, it is
\begin{equation}
    \sigma^{(2,0)} = \frac13\sum_{a} \Big< \frac1{r_a}\Big>\,,
\end{equation}
where $r_a$ is the distance between the nucleus and the $a$'th electron, and the summation runs over all electrons.
The expectation value is taken with the nonrelativistic wave function in the limit of an infinitely heavy nucleus.

The next most important correction in the light H$^-$ system is the leading nuclear mass correction
\begin{align}
\label{sigma21}
\sigma^{(2,1)} \equiv &\, \frac{m}{M} \left(\frac{1-g_N}{g_N}\,\sigma^{(2)}_A + \sigma^{(2)}_B\right)
 \nonumber \\ = &\,
 \frac{m}{M} \left(\frac{1-g_N}{g_N}\,\frac{ \left< p_N^2 \right> }{3Z}\,
 + \frac13 \Big< \sum_a \frac1{r_a} \frac1{(E-H)'} p_N^2\Big>\right.
 \nonumber \\ &
 + \left.\frac13 \left< \sum_a \vec{r}_a \times \vec{p}_N \frac1{(E-H)}
                 \sum_b \frac{\vec{r}_b}{r_b^3} \times \vec{p}_b\right> \right)\,,
\end{align}
where $1/(E-H)$ is the Green function and
$1/(E-H)'$ is the reduced Green function (with the reference state removed from the sum over the spectrum), 
$\vec{p}_N = -\sum_a\vec{p}_a$, where $\vec{p}_a$ is the momentum of the $a$'th electron, $g_N = (M/m_p)(\mu/\mu_N)/(ZI)$
is the nuclear $g$-factor;
$M$, $I$ and $\mu$ are the nuclear mass, spin and the magnetic moment,
respectively; $m_p$ is the proton mass and $\mu_N$ is the nuclear magneton.

The relativistic correction to nuclear magnetic shielding 
(of order $\alpha^4$) was derived in Ref.~\cite{Rudzinski2009_JCP}.
The obtained formula is
\begin{widetext}
\begin{align}\label{eq:sigma4}
\sigma^{(4,0)} &= \sigma_1^{(4)} + \sigma_{2A}^{(4)} + \sigma_{2B}^{(4)} +
\sigma_{2C}^{(4)} + \sigma_3^{(4)} \nonumber \\
&= 
\frac{2}{3}\,\biggl\langle\biggl(\frac{1}{r_1}+\frac{1}{r_2}\biggr)\,
\frac{1}{(E-H)'}\,\biggl[\sum_a \biggl(\frac{\pi\,Z}{2}\,\delta(\vec r_a)-\frac{p_a^4}{8}\biggr)
+\pi\,\delta(\vec r)-\frac{1}{2}\,p_1^i\biggl(\frac{\delta^{ij}}{r}+\frac{r^i\,r^j}{r^2}\biggr)\,p_2^j
\biggr]\biggr\rangle
 \nonumber \\
& -\frac{2}{9}\,\biggl\langle\pi\,\bigl[\delta(\vec
r_1)-\delta(\vec r_2)\bigr]\,
\frac{1}{(E-H)}\,\biggl[
3\,p_1^2-3\,p_2^2-\frac{Z}{r_1}+\frac{Z}{r_2}-\frac{\vec r\cdot(\vec r_1+\vec r_2)}{r^3}
\biggr]\biggr\rangle
 \nonumber \\ 
&-\frac{1}{6}\,\biggl\langle\biggl(
\frac{\vec r_1\times\vec p_1}{r_1^3} + \frac{\vec r_2\times\vec p_2}{r_2^3}\biggr)\,
\frac{1}{(E-H)}\biggl[
\vec r_1\times\vec p_1\,p_1^2 + \vec r_2\times\vec p_2\,p_2^2
+\frac{1}{r}\,\vec r_1\times\vec p_2 + \frac{1}{r}\,\vec r_2\times\vec p_1
-\vec r_1\times\vec r_2\,\frac{\vec r}{r^3}\cdot(\vec p_1+\vec p_2)\biggr]\biggr\rangle
 \nonumber \\
&-\frac{1}{8}\,\biggl\langle
\biggl(\frac{r_1^i\,r_1^j}{r_1^5}-\frac{r_2^i\,r_2^j}{r_2^5}\biggr)^{(2)}\,
\frac{1}{(E-H)}\,\biggl(Z\,\frac{r_1^i\,r_1^j}{r_1^3} - Z\,\frac{r_2^i\,r_2^j}{r_2^3}
+\frac{r^i}{r^3}\,(r_1^j+r_2^j)\biggr)^{(2)}\biggr\rangle \\
&+ \sum_a
\biggl\langle
\frac{1}{12\, r_a^3}\,
\Big(\frac{\vec r\cdot\vec r_1\;\vec r\cdot\vec r_2}{r^3} - 3\,\frac{\vec r_1\cdot\vec r_2}{r}\Big)
-\frac{1}{6}\,
 \Big( \frac{1}{r_a}\,p_a^2+\frac{(\vec r_a\times\vec p_a)^2}{r_a^3}+4\,\pi\,\delta(\vec r_a)
 \Big)
\biggl\rangle \nonumber 
\,,
\end{align}
\end{widetext}
where $\vec r\equiv \vec r_1-\vec r_2\,$, $(p^i\,q^i)^{(2)} =
p^i\,q^j/2+p^j\,q^i/2-\delta^{ij}\,\vec p\cdot\vec q/3$. Following Ref. \cite{Rudzinski2009_JCP} we split the relativistic correction into five terms according to the form of the matrix elements. The first term, $\sigma_1^{(4)}$, has intermediate states of $^1\textrm{S}$ symmetry. The second $\sigma_{2A}^{(4)}$, the third $\sigma_{2B}^{(4)}$ and the fourth $\sigma_{2C}^{(4)}$ terms involve the intermediate states of the symmetry $^3\textrm{S}$, $^1\textrm{P}^e$, and $^3\textrm{D}$, respectively. The final part, $\sigma_3^{(4)}$, contains the first-order matrix elements.

The logarithmic part of the QED correction to the magnetic shielding was first derived in~\cite{Rudzinski2009_JCP}, 
and a full leading-order QED correction for low-Z systems was later derived in Ref.~\cite{Wehrli2022_PRA}. 
Since the calculation of the full QED correction to nuclear magnetic shielding is very demanding numerically~\cite{Wehrli2021_PRL}, 
as it requires calculation of the perturbed Bethe-type logarithms, we directly calculate only the logarithmic part of the QED correction,
while the nonlogarithmic part is estimated from the atomic hydrogen value.
The logarithmic part of the QED correction is given by the formula 
\begin{align}
\sigma^{(5)}_{\rm log} = &\,
 \ln (Za)\,\bigg[
   -\frac{16 Z}{9}\, \Big< \sum_a\frac1{r_a}\, \frac1{(E-H)'}\, \sum_b\delta(\vec{r}_b)\Big>
\nonumber \\ 
   + & \frac{28}{9}\, \Big< \sum_a\frac1{r_a}\, \frac1{(E-H)'}\, \delta(\vec{r})\Big>
   - \frac{40}{9}\, \Big< \sum_a\delta(\vec{r}_a)\Big>
   \bigg]
\end{align}
from Ref.~\cite{Yerokhin2024_PRA}.

\section{Computational methods}
Our computations are carried out in the basis of the exponential-type functions following the approach of Korobov~\cite{Korobov2000, Korobov2002}. For $^1\textrm{S}$ state, these functions have the form
\begin{equation}
\label{eq::exp_function}
\phi(r_1,r_2,r) = \sum_{i=1}^{\cal N} v_i
[e^{-\alpha_i r_1-\beta_i r_2-\gamma_i r} \pm (r_1 \leftrightarrow r_2)],
\end{equation}
where $\alpha_i$, $\beta_i$, and $\gamma_i$ variational parameters are randomly generated from the quasi-random distribution with the following conditions
\begin{eqnarray}
A_1<\alpha_i<A_2,\;\; \beta_i+\gamma_i>\varepsilon, \nonumber \\
B_1<\beta_i<B_2,\;\; \alpha_i+\gamma_i>\varepsilon, \\
C_1<\gamma_i<C_2,\;\; \alpha_i+\beta_i>\varepsilon, \nonumber
\end{eqnarray}
with $\varepsilon = 0.2$, approximately equal to $\sqrt{2\,m\,(E_{\mathrm{H}}-E_{\mathrm{H}^-})}$. In order to improve the accuracy of the wave function, we used a doubled set of parameters $A_i,B_i,C_i$. The values of these two sets of parameters were determined by minimizing the non-relativistic energy using the variational method. The coefficients of the linear combinations -- $v_i$ -- were determined by solving the generalized eigenvalue problem. In order to reach the desired numerical accuracy, calculations were performed in quad-double precision arithmetic (with about 62 decimal digits). 

All matrix elements are expressed in terms of rational, logarithmic, and dilogarithmic functions in $\alpha,\beta$, and $\gamma$~\cite{Korobov2002a},
so they can be calculated very efficiently.
The size of the base set starting with 300 functions was increased by 300 functions in each step 
to reach up to 1800 basis functions, yielding a non-relativistic energy of the ground $^1\mathrm{S}$ state of the H$^-$ ion
\begin{equation}
    E_0(1^1S_0) = -0.527~751~016~544~377~196
\end{equation} 
which agrees with even more accurate results~\cite{Nakatsuji2007,Korobov2018,Drake2025} up to all digits shown

While the convergence of the non-relativistic energy and the first-order matrix elements with the size of the basis set is very fast, 
the accurate calculation of the second-order matrix elements is numerically much more challenging. 
The inversion of the operator $E-H$ is performed in a basis set of even parity functions with $l=0,1,2$ of the form given in Eq. (\ref{eq::exp_function}) as well as
\begin{equation}
\vec{\phi}(r_1,r_2,r) = \sum_k v_k\,\vec r_1 \times \vec r_2\, [e^{-\alpha_k r_1-\beta_k r_2-\gamma_k r} - (r_1 \leftrightarrow r_2)]\,,    
\end{equation}
and
\begin{align}
    \phi^{ij}(r_1,r_2,r) & = \sum_k v_k\, [(r_1^i\,r_1^j-r_1^2\,\delta^{ij}/3)\ \nonumber \\
& \times e^{-\alpha_k r_1-\beta_k r_2-\gamma_k r}- (r_1 \leftrightarrow r_2)] \nonumber
\\ & +\sum_l v_l\, (r^i\,r^j-r^2\,\delta^{ij}/3) \nonumber \\
& \times [e^{-\alpha_l r_1-\beta_l r_2-\gamma_l r}- (r_1 \leftrightarrow r_2)] 
\end{align}
respectively. The values of the parameters $A_i$, $B_i$ and $C_i$ for the functions above are obtained by variational optimization of the functionals which are the symmetric second-order matrix elements with the corresponding operator standing on the right-hand side of the resolvent operators in the Equations \eqref{sigma21} and \eqref{eq:sigma4}. The number of basis functions in the $E-H$ operator of the second-order matrix elements is doubled with respect to the number of basis functions used in the calculations of the energy and first-order matrix elements (starting with 600 and increasing by 600 functions up to 3600 basis functions). 
Optimization of the perturbed wave functions is the most computationally demanding step of the presented calculations.

\section{Results and discussion}

The obtained numerical values of the expectation values and contributions to the nuclear magnetic shielding of H$^-$ calculated with the largest basis (of 1800 functions) are presented in Table \ref{tab:shieldingpartial}. All digits in that table are significant. The main source of the numerical uncertainty in our calculations comes from the second-order matrix element $\sigma_{2C}^{(4)}$. This matrix element is highly singular and converges with the size of the basis much more slowly than all the other matrix elements. 
Fortunately, its magnitude is two orders smaller than the leading contributions to $\sigma^{(4)}$,
which makes the absolute numerical accuracy sufficiently high. Moreover,
uncertainties coming from higher-order expansion terms such as $\sigma^{(2,2)}$, $\sigma^{(4,1)}$ and $\sigma^{(5)}$ 
dominate the overall uncertainty budget of the present calculations.

\renewcommand{\arraystretch}{1.3}
\begin{table}[b]
\caption{Numerical values for partial contributions to the nuclear magnetic shielding in $^1$H$^-$. All digits are significant. The values are given in atomic units and the contributions to $10^6\sigma$ are given by $10^6\alpha^n\,\sigma^{(n)}$.}
\label{tab:shieldingpartial}
\begin{ruledtabular}
\begin{tabular}{ld{1.10}d{2.10}}
\centt{operator} & \centt{expectation value} & \centt{contribution to $10^6\sigma$} \\ 
\hline
$\sigma^{(2,0)}$               &  0.455\,507\,845\,10 & 24.256\,409\,7 \\
$\sigma^{(2)}_A$             &  0.373\,754          & -0.008\,898\,9 \\
$\sigma^{(2)}_B$             & -0.535\,115          & -0.015\,519\,2 \\
$\sigma^{(4)}_1$             &  1.840\,204          & 0.005\,218\,3  \\
$\sigma^{(4)}_{2A}$          &  3.172\,855          & 0.008\,997\,3  \\
$\sigma^{(4)}_{2B}$          & -0.001\,894          & -0.000\,005\,4  \\
$\sigma^{(4)}_{2C}$          &  0.154\,32(36)       & 0.000\,437\,6(11) \\
$\sigma^{(4)}_3$             & -1.230\,166          & -0.003\,488\,4 \\
$\sigma^{(5)}_{\log}$        &  2.283\,049          &  0.000\,047\,2 \\
\end{tabular}
\end{ruledtabular}
\end{table}

The partial contributions to the shielding constant presented in Table \ref{tab:shieldingpartial} reveal several unique features of the magnetic shielding in the H$^-$ system. First, the leading-order finite-mass corrections are more than twice as large as the leading-order relativistic corrections and contribute to the magnetic shielding constant at the level of 0.1\%. 
This highlights the importance of using the quantum-electrodynamic theory to derive the finite-mass effects for accurate calculations of the relativistic properties of atomic and molecular systems, such as magnetic shielding. 
Due to the use of perturbative expansion in the nuclear mass, the magnetic shielding constants for deuteride and tritide may be obtained with the formula given in Equation \eqref{sigma21} using expectation values $\sigma^{(2)}_A$ and $\sigma^{(2)}_B$ given in Table \ref{tab:shieldingpartial} with appropriate nuclear masses and $g$-factors. 

Regarding the relativistic correction to the magnetic shielding, the largest contribution comes from the second-order matrix element $\sigma^{(4)}_{2A}$, 
which mixes singlet and triplet states. 
H$^-$ has only one bound state~\cite{Hill1977_PRL, Hill1997_JMP}, so in this system this contribution comes entirely from mixing with the continuum states. The antiscreening effect reported in studies of helium \cite{Rudzinski2009_JCP, Yerokhin2024_PRA} and helium-like cations is particularly pronounced in the hydrogen anion. In this case, the total relativistic correction to shielding in H$^-$ system is more than twice as large as the hydrogenic value. 
This supports the explanation of the antiscreening effect given in Ref.~\cite{Yerokhin2024_PRA}.

The final results for nuclear magnetic shielding of $^1$H$^-$ along with the previously calculated values for $^1$H, $^3$He$^+$ and $^3$He systems are given in Table \ref{tab:shieldingstotals}. 
The higher order contributions for H$^-$ are estimated as follows. The second-order finite mass correction $\sigma^{(2,2)}$ of H$^-$ is estimated from the ratio of $\sigma^{(2,2)}$ to $\sigma^{(2,1)}$ in atomic hydrogen multiplied by the calculated value of $\sigma^{(2,1)}$ for H$^-$.
The uncertainty of this estimation is set to 25\% of the $\sigma^{(2,2)}$ value.
Since the calculations of the full $\sigma^{(5,0)}$ term in $^3$He in Ref.~\cite{Wehrli2021_PRL,Wehrli2022_PRA} had shown significant cancellation between the logarithmic and non-logarithmic part of the leading-order QED correction, instead of using the calculated $\sigma^{(5)}_{\mathrm{log}}$ value, we use the hydrogenic value of $\sigma^{(5,0)}$ and assign 25\% uncertainty of $\sigma^{(5,0)}$ to this estimation.  Since $\sigma^{(4,1)}$ is unknown, we estimate its uncertainty as $100\%$ of $\alpha^2\sigma^{(2,1)}$, because
for H$^-$ it is larger than $\tfrac{m}{M} \sigma^{(4,0)}$. Finally, $\sigma^{(6,0)}$ is estimated by taking twice the hydrogenic value from Ref.~\cite{Pachucki2023_PRA},
which includes only the known relativistic contribution to $\sigma^{(6,0)}$, and $\sigma_\text{fs}$ is taken to be the same as in the case of atomic hydrogen. The summary of the contributions to the final value of $\sigma$ in $^1$H$^-$ along with the assigned uncertainties is given in Table~\ref{tab:shieldingstotals}. The main source of the uncertainty in the present calculation is the unknown relativistic recoil correction to the magnetic shielding $\sigma^{(4,1)}$. Compared with previously calculated values of the magnetic shielding in H$^-$~\cite{Boucekkine1988,Vaara2003} (shown in Table \ref{tab:shieldingstotals}), present results offer four orders of magnitude improvement in the accuracy of the results. Additionally, in Table~\ref{tab:shieldingstotals} we give the updated the value of the shielding constant of the $^3$He taking into account the improved accuracy of the $\sigma^{(4,0)}$(He) given in Ref.~\cite{Yerokhin2024_PRA}. However, this change is only $7 \times 10^{-12}$ so it is well within the estimated uncertainty of $\sigma$ in $^3$He which is $24 \times 10^{-12}$.

It is also worth noting that  the quadratic terms in the magnetic field $\sim \vec{B}^2$ may contribute to the magnetic shielding of the nucleus. Its value for H$^-$ was determined to be $21.0208 \times 10^{-15}\,\textrm{T}^{-2} $ in Ref.~\cite{Vaara2003}, which is negligible at the current precision level with laboratory $B$-fields.

\newcommand{\mct}[1]{\multicolumn{2}{l}{#1}}
\begin{table*}[t]
\caption{Contributions to the shielding constant $10^6\sigma$ for $^1$H, $^1$H$^-$, $^3$He$^+$ and $^3$He. Results for $^1$H$^-$ are all new while the results for $^1$H, $^3$He$^+$ and $^3$He are taken from Ref.~\cite{Pachucki2023_PRA} with the exception of the more accurate $\sigma^{(4,0)}$(He) taken from \cite{Yerokhin2024_PRA}. All values are given in atomic units.}
\label{tab:shieldingstotals}
\begin{ruledtabular}
\begin{tabular}{ld{2.12}d{2.12}d{2.12}d{2.12}}
 & \centt{$^1$H} &  \centt{$^1$H$^-$} & \centt{$^3$He$^+$} & \centt{$^3$He} \\ 
\hline
$\alpha^2\,\sigma^{(2,0)}$ &17.750\,451\,5 & 24.256\,410\,7 & 35.500\,903\,0 &  59.936\,771\,0  \\
$\alpha^2\,\sigma^{(2,1)}$ & -0.017\,603\,7& -0.024\,418\,1 & -0.013\,933\,4 &  -0.023\,020\,1  \\
$\alpha^2\,\sigma^{(2,2)}$ &  0.000\,014\,1& 0.000\,019\,6(49) & 0.000\,001\,4& 0.000\,002\,1(7) \\
$\alpha^4\,\sigma^{(4,0)}$ &0.002\,546\,9 & 0.011\,159\,4(11) &  0.020\,375\,1 & 0.052\,670\,7 \\
$\alpha^4\,\sigma^{(4,1)}$ &  0.000\,000\,0(28) & 0.000\,000\,0(60) & 0.000\,000\,0(74)& 0.000\,000\,0(192)   \\
$\alpha^5\,\sigma^{(5,0)}$ & 0.000\,018\,4 & 0.000\,018\,4(46) &  0.000\,082\,0 &   0.000\,096\,3\\
$\alpha^6\,\sigma^{(6,0)}$ & 0.000\,000\,2(2) & 0.000\,000\,4(4) & 0.000\,006\,5(65)  & 0.000\,012\,9(129)  \\
$\sigma_\mathrm{fs}$ & -0.000\,000\,1 & -0.000\,000\,1(1) & -0.000\,006\,7  & -0.000\,013\,5(67)  \\[2ex]
$10^6\,\sigma$& 17.735\,427(3) & 24.243\,189(9) & 35.507\,427(10) &   59.966\,519(24)\\
$10^6\,\sigma$, RHF \cite{Boucekkine1988}  & & 24.21 &  & \\
$10^6\,\sigma$, Full CI \cite{Vaara2003} & & 24.2305 &  & 59.8925 \\
\end{tabular}
\end{ruledtabular}
\end{table*}

\section{Summary}

We have presented a highly accurate calculation of the nuclear magnetic shielding in the hydrogen anion using the NRQED theory. The magnetic shielding constant is obtained with an accuracy of nine-parts-per-trillion (9 ppt) by rigorously calculating finite nuclear mass and relativistic effects with high numerical accuracy, as well as by estimating the effects of the higher-order corrections. Inclusion of these effects has allowed us to obtain the shielding constant with the relative uncertainty of 0.37 ppm. The final value of the constant of $^1$H$^-$ is $\sigma = 24.243\,189(9) \times 10^{-6}$.
This result will allow for a stringent test of the NRQED theory of magnetic shielding as well as for verification of CPT symmetry by comparison of proton-to-antiproton magnetic moments.

Additionally, this calculation illustrates the importance of the first-principle theoretical approach
for the effects of finite nuclear mass effects on the magnetic properties of light atoms. 
Obtained results may be used to test and validate common quantum-chemical methods
for the calculation of magnetic properties of atoms and molecules by testing those methods 
on the simple case of the nuclear magnetic shielding of H$^-$.
\newline

\begin{acknowledgments}
This research was supported by the National Science Center (Poland) Grant No.
2018/29/N/ST4/02034.
\end{acknowledgments}

\end{document}